\documentclass[aps,pra,showpacs,twoside,twocolumn,10pt]{revtex4-2}
\usepackage[colorlinks=true, citecolor=red, urlcolor=blue ]{hyperref}
\usepackage{epsfig,amssymb,amsfonts,amsmath,bm,subfigure,mathtools,amsthm,braket,soul,enumitem,xcolor,physics,graphics,graphicx}
\UseRawInputEncoding
\usepackage[normalem]{ulem}
\usepackage{comment}
\usepackage{epstopdf}
\usepackage{multirow}
\usepackage{cancel}
\usepackage{soul}

\newtheorem{proposition}{Proposition}

\newcommand{\added}[1]{{\color{black}#1}}

\begin{document}

\title{
Recovery of resources through sequential noisy measurements
}
\author{Sudipta Mondal$^{1}$, Pritam Halder$^{1}$, Amit Kumar Pal$^{2}$, Aditi Sen (De)$^{1}$}
\affiliation{$^1$Harish-Chandra Research Institute, A CI of Homi Bhabha National Institute, Chhatnag Road, Jhunsi, Prayagraj 211 019, India}
\affiliation{$^2$ Department of Physics, Indian Institute of Technology Palakkad, Palakkad 678 623, India}

\begin{abstract}

Noisy unsharp measurements incorporated in quantum information protocols may hinder performance, reducing the quantum advantage.  However, we show that, unlike projective measurements which completely destroy quantum correlations between nodes in quantum networks, sequential applications of noisy measurements can mitigate the adverse impact of noise in the measurement device on quantum information processing tasks.  We  demonstrate this in the case of concentrating entanglement on chosen nodes in quantum networks via noisy measurements performed by assisting qubits. In the case of networks with a cluster of three or higher number of qubits, we exhibit that sequentially performing optimal unsharp measurements on the assisting qubits yields localizable entanglement between two nodes akin to that obtained by optimal projective measurements on the same assisting qubits.  Furthermore, we find that the proposed approach using consecutive noisy measurements can potentially be used to prepare desired  states that are resource  for specific quantum schemes. We also argue that assisting qubits have greater control over the qubits on which entanglement is concentrated via unsharp measurements, in contrast to sharp measurement-based protocols, which may have implications for secure quantum communication.

\end{abstract}

\maketitle

\section{Introduction}
\label{sec:intro}

\begin{figure}
    \centering
    \includegraphics[width=\linewidth]{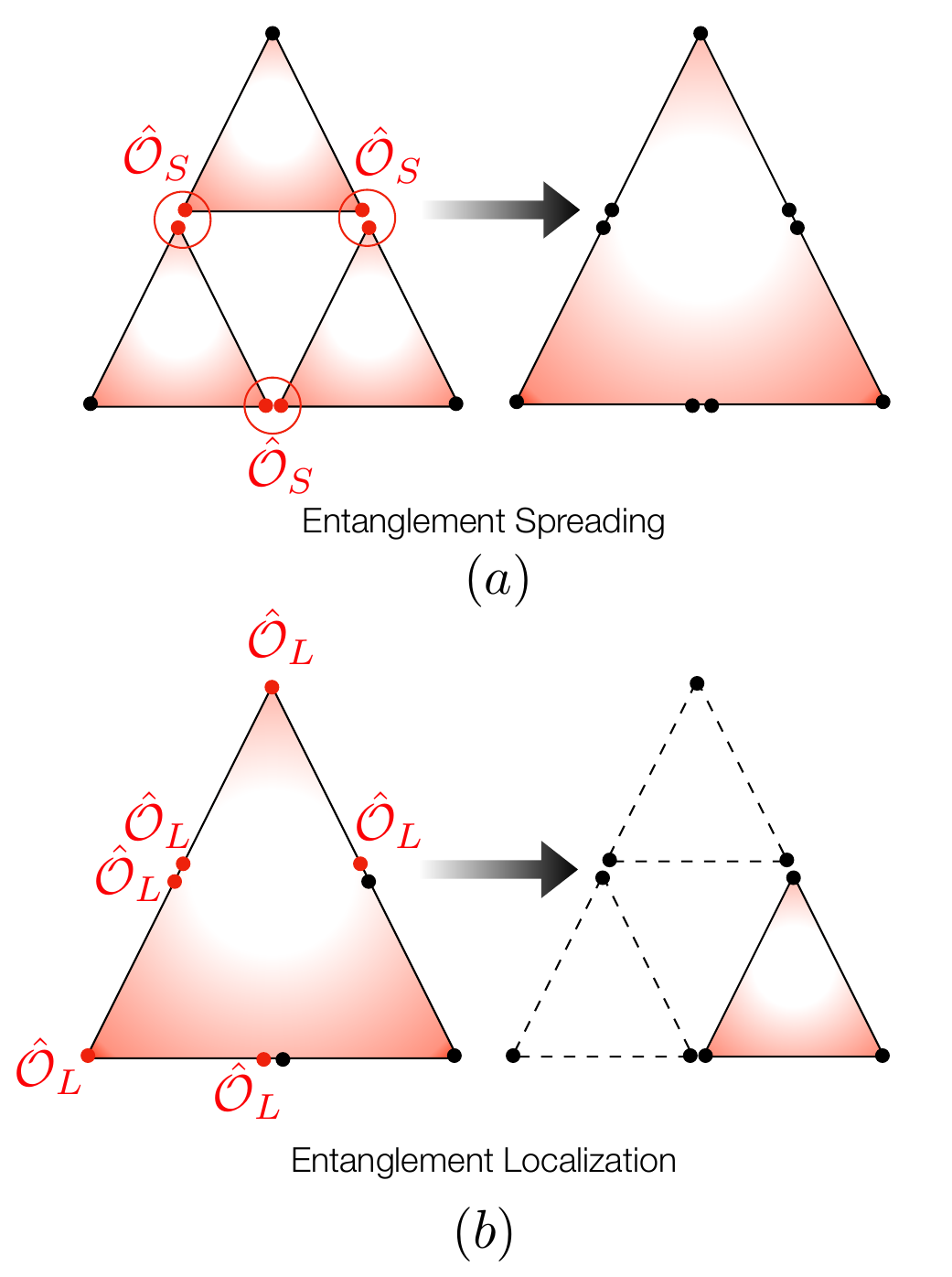}
    \caption{\textbf{Spreading and localizing entanglement on a network.} (a) Three small three-qubit clusters, each hosting an entangled state, can be merged together via application of two-qubit operations $\hat{\mathcal{O}}_S$ -- either a two-qubit entangling unitary, or a measurement of two qubits -- to form an \emph{entangled} $9$-qubit state. This can be imagined as spreading of entanglement over a network of nine qubits. (b) Starting from a network of nine qubits sharing an entangled state, one can perform operations $\hat{\mathcal{O}}_L$ on selected qubits such that entangled states can be created on three chosen qubits. This can be seen as entanglement localization, where the operations $\hat{\mathcal{O}}_L$ are typically projection measurements. We point out that the complete destruction of links between assisting and non-assisting parties can be avoided if the measurements are noisy.  Hence using the assisting qubits repeatedly to share  certain entangled states exploiting noisy measurements provides more control on the protocol than the one with PV measurements.  }
    \label{fig:network}
\end{figure}

Establishing large-scale quantum networks with multiple nodes connected by entangled quantum channels \cite{Kimble2008,wehner_18internet,Pirker2019,Wei2022,azuma2023quantum} is a crucial endeavor for realizing several information processing tasks, such as distributed quantum computing \cite{Beals_2013}, quantum sensing \cite{Giovannetti2011}, and quantum communication protocols \cite{Bennett_1992,Bennett_1993,Mattle_1996, Bouwmeester_1997, Murao_1999, pati_2000, grudka_2004,Bruss_2004,Bennett_2005,    BRU__2006,Sen(de)_2010, de_2011, Horodecki_2012, Shadman_2012,Das_2014} including the secure ones \cite{Ekert_1991, Hillery_1999, Shor_2000, Adhikari_2010, Bennett_2014, Sazim_2015, Ray_2016, mudholkar_2023}. On one hand, quantum networks are desired to distribute entanglement, or other resources, over two, or multiple nodes, even at large distances~\cite{horodecki2009}. On the other hand, it is also essential to prepare specific resourceful states among a small number of nodes in a large network, and subsequently isolate them from the rest of the network, for executing specific quantum protocols over all scales in the quantum network. While the focus of the former enterprises has been the spreading of entanglement over a quantum network~\cite{ZeilingerHorne1997,briegel1998,dur1999,WaltherZeilinger2005,Browne2005,acin2007,Tashima(3)2008,Tashima(1)2009,Tashima(2)2009,Tashima(4)2011,Ozdemir_2011,Bugu2013,Ozaydin2014,zang2015,Severin2021,inflation2021,circulation2022}, the latter \emph{concentrates} entanglement over a small subset of nodes (see Fig.~\ref{fig:network}) by performing local projective (PV) measurements at the single-qubit level, which are often easy to carry out in experiments. The corresponding protocol is referred to as \emph{localizing} entanglement~\cite{divincenzo1998, smolin2005, popp2005, gour2006} and the qubits on which PV measurements are carried out are referred to as assisting qubits. 

\added{In a multi-qubit system represented by the state $\rho$, the \emph{localizable entanglement} is the maximum average entanglement that can be concentrated in a subset $A$ of qubits by performing measurements on the rest of the qubits, forming the subset $B$. Mathematically, it is given by 
\begin{eqnarray}
    \mathcal{E}(\rho) = \max \sum_\lambda p_\lambda E\left(\Tr_B[\rho_\lambda]\right),
\end{eqnarray}
where $\rho_\lambda$ is the post-measurement state corresponding to the measurement outcome $\lambda$, occurring with probability $p_\lambda$ ($\sum_\lambda p_\lambda = 1$), and $E$ is a pre-decided entanglement measure~\cite{divincenzo1998, smolin2005, popp2005, gour2006}.
Here, the maximization is performed over all possible local projective measurements on all qubits in the subsystem $B$. Along with investigation of the properties of localizable entanglement~\cite{sadhukhan2017,Pollock2021,Krishnan2023,vairogs2024}, the idea has been used to characterize entanglement in subsystems of the Greenberger Horne Zeilinger (GHZ) and the stabilizer states \cite{hein2006entanglementgraphstatesapplications, Amaro2018, Amaro2020}, to define correlation length in one-dimensional quantum spin models \cite{PhysRevLett.92.027901, PhysRevLett.92.087201, PhysRevA.71.042306, PhysRevA.69.062314}, to characterize quantum phase transitions in the cluster-Ising \cite{PhysRevA.80.022316, PhysRevA.84.022304} and cluster-XY models \cite{PhysRevE.86.021101} as well as locally perturbed topological quantum codes~\cite{HK2022,HK2025},  and in the study of entanglement percolation in quantum networks~\cite{nat_phy_per}. 
Moreover, the idea has been shown to be closely entwined with realizing  quantum gates in measurement-based quantum computing~\cite{mbqc_Raussendorf2003}, and with identifying resource in controlled quantum communication protocols~\cite{Barasiński2018}}. 

One of the inevitable challenges in the construction of a quantum network is decoherence \cite{Zurek03, breuer2002}. The noise can enter the network  either through  single and two-qubit quantum gate operations \cite{Lidardyndecoup,McClean23errormitigation,mondal2023imperfect} or at the time of  measurements \cite{Busch_2013} used  for  spreading \cite{inflation2021, mondal2023duality}.  In all these situations, resources like entanglement~\cite{horodecki2009}, discord~\cite{Bera_2017}, and coherence~\cite{Streltsov_2017} get affected, resulting in a degradation of the prepared quantum state to be used as resource. Over the years, several strategies are developed to protect, and correct quantum states during the preparation of quantum states. Along with the development of  quantum error correcting strategies~\cite{Steane_2006,nielsenchuang}, recent studies have also addressed the possibility of preserving quantum correlations against noise by preparing certain class of quantum states or by tuning parameters of the evolving Hamiltonian. In particular, necessary and sufficient conditions for freezing  of quantum correlations like quantum discord and work-deficit~\cite{Mazzola2010,chanda2015} and preventing sudden death of entanglement~\cite{yu2009} are derived in the case of bipartite as well as multipartite states subjected to local noisy channels. It has also been shown that entanglement in quantum many-body systems, affected by local Markovian noise, can  be preserved over time by  properly choosing initial conditions~\cite{Carnio15, chanda2018}.  

Decoherence can influence entanglement localization protocol in two different ways -- (a) it can either affect the multipartite initial state used for localizing entanglement~\cite{Amaro2018, Amaro2020,Banerjee_2020,Banerjee2022}, or (b) the measurement operators applied on assisting parties can be noisy (or \emph{unsharp}) due to the interaction of the measurement apparatus with the environment~\cite{Busch_2013}. In this work, we will focus on the consequence of the latter. Similar to the favourable impact of the non-Markovian noise over the Markovian one exhibited through the revival of entanglement after collapse~\cite{Mazzola2009,Fanchini2017,Gupta2022}, noisy measurements have the advantage of retaining quantum correlations between the measured and the unmeasured parties, thereby allowing repeated applications of the noisy measurement on the same party. Such sequential measurements are used to verify entanglement, or nonlocal correlations, between two or more observers in different times  with the help of Bell inequality \cite{Mal16,Silva15,asmita19,Colbeck20}, bilocal inequality \cite{Halder2022}, entanglement witness \cite{Bera18,Srivastava21}, contextuality \cite{Anwer2021}, and steering~\cite{sasmal18, Shenoy19} to name a few, along with advantage in communication tasks including teleportation~\cite{ROY2021127143} and telecloning~\cite{Das2023}. In this paper, we explore whether {\it sequential}  noisy measurements on assisting nodes can overcome the effects of measurement-noise in concentrating entanglement over  subsystems in a quantum network, and answer the question affirmatively.  

Towards this, we formulate the definition of localizable entanglement via multiple rounds of noisy measurements on a selected subset of assisting qubits, where white noise is assumed to be present in the measurement apparatus. We demonstrate an equivalence between  the optimization involved in localizable entanglement over all possible  directions in all rounds of noisy measurements, and  the optimization performed sequentially for each round of measurement, which subsequently reduces the complexity in the optimization. We use this to show, for three-qubit generalized Greenberger Horne Zeilinger (gGHZ) \cite{Greenberger2007} and generalized W (gW) \cite{Durvidal2002} states, that for moderate noise strengths, the LE on any two qubits via sharp projection measurements on the assisting third qubit can be achieved up to an negligible error (of $5\times 10^{-3}$) through four to six rounds of noisy measurements. This result remains unchanged for three-qubit pure states of W class \cite{Durvidal2002,Yang2009}, while for states from the GHZ class \cite{Durvidal2002,Yang2009}, LE increases very slowly with increasing the rounds of noisy measurements, indicating a much larger number of measurement rounds for achieving it corresponding to projection measurements. Further,  we observe that for a fixed assisting qubit,  the favourable noisy measurement directions in two consecutive rounds  can be orthogonal. Beyond three qubits, we find that  the LE corresponding to sharp projection measurements is almost equal to the one obtained via six rounds of noisy measurements for the generalized GHZ states \cite{Greenberger2007} and the Dicke states \cite{dicke1954,bergmann2013,lucke2014,KUMAR20171701} with moderate number of qubits. However, our numerical search illustrates that with increasing number of parties, more number of rounds are required to concentrate the entanglement that is localizable  through PV measurements. Moreover, we  point out specific patterns in the optimal measurement directions for multiple sequential rounds of measurements that lead to the localization of entanglement equal to the same corresponding to sharp  measurements. Further, for three-qubit
GHZ and W states, we demonstrate that multiple rounds of unsharp measurements effectively facilitate the creation of high-fidelity maximally entangled states with high probability which typically happens with PV measurement.

We further realize that the noisy measurements can provide an additional controlling power to the assisting qubits during concentration of entanglement, or quantum communication scheme. Consider a network having \(N = N_A + N_B\) qubits, in which a quantum protocol, $\mathbb{P}$, to be implemented requires $E_A$ amount of entanglement over the subsystem $A$ of $N_A$ qubits.  This is possible when single-qubit projective measurements on all  \(N_B\) assisting  qubits in the subsystem $B$ are performed, thereby destroying the links between the members of all qubit-pairs --  one qubit belonging to $A$ while the other is in $B$. Note that the assisting qubits in $B$ cannot achieve the goal of localizing $\mathcal{E}_A$ amount of entanglement in the \emph{first round} of \emph{noisy} measurements, and unlike PV measurements, they are still entangled with the qubits in $A$. Hence, they  keep on performing a total of, say, $R$ rounds of noisy measurements till they concentrate $\mathcal{E}_A$ amount of entanglement on $A$. This is advantageous in scenarios where parties in $A$ are not fully trustworthy, as one or a group of the assisting parties, upon finding the malign intention of one or more parties in $A$ after, say,  $R_c$ ($<R$) rounds of measurements,  may stop performing  measurements so that  the parties in $A$ fall short of localizing $\mathcal{E}_A$ amount of entanglement, thereby failing to execute the protocol $\mathbb{P}$ successfully due to shortage of resource. In this way, requirement of a higher number of rounds of noisy measurements to localize $E_A$ amount of localizable entanglement over $A$  provides more time to the assisting parties to decide, or verify, the reliability of the parties in $B$, thereby ensuring a finer control over the localization strategy. Our results on localizing $\mathcal{E}_A$ amount of entanglement on the subsystem $A$ via multiple rounds of noisy measurements, as discussed in the subsequent sections, demonstrate that this control can indeed be achieved.

The organization of the paper is as follows. The architecture of networks in which unsharp measurements are performed sequentially is described in Sec. \ref{sec:prelim}. In Sec. \ref{sec:three_qubit}, the beneficial role of sequential noisy measurements on localizable entanglement is illustrated when the network is composed of several clusters containing three qubits. Further, the use of the multiple round measurement protocol for preparing  desired resourceful state is demonstrated with three-qubit GHZ and W states in Sec. \ref{sec:preparation}. Sec. \ref{sec:multiple_qubits} discusses the extension of the results to cluster of multiple qubits, described by generalized GHZ and the Dicke state, where entanglement is localized over two qubits via noisy measurements of multiple rounds on all of the rest of the qubits. The concluding remarks and outlook are included in Sec. \ref{sec:conclude}.

\section{Framework of localizable entanglement with noisy measurements}
\label{sec:prelim}

Let us mathematically set up the problem of localizing (concentrating) entanglement over a subsystem of a multiqubit quantum network via unsharp measurements on the rest of the qubits. In this framework, we assume that the measurement apparatus is no more isolated, but is in contact with environment(s), thereby giving rise to the possibility of noisy measurements during localizing  entanglement. 



\subsection{Localizing entanglement via unsharp measurements}
\label{subsec:single_round}

Let $A:B$ be a bipartition of a $N$-qubit quantum network in the state $\rho$, where the qubits in the partition  $A$ ($B$) of size $N_A$ ($N_B$) are labeled by $a$ ($b$) with  $a=1,2,\cdots,N_A$ ($b=1,2,\cdots,N_B$), and $N_A+N_B=N$. To implement certain tasks, only $N_A$ qubits are required and hence single-qubit projection measurements are performed on all qubits, in $B$, calling them as the \emph{assisting qubits},  such that non-vanishing post-measured average entanglement is localized post-measurement over the qubits in $A$. \added{We consider a \emph{sharp} single-qubit projective measurement on the qubit $b\in B$ to be denoted by 
$P(\lambda_b,\hat{n}_b)=\frac{1}{2}[I+\lambda_b \hat{n}_b.\vec{\sigma_b}]$, where $I$ is the identity operator on the qubit Hilbert space, $\vec{\sigma}_b=\left(\sigma^x_b,\sigma^y_b,\sigma^z_b\right)$ are the Pauli matrices, and  $\lambda_b=\pm 1$ are the measurement outcome with $\sum_{\lambda_b}P(\lambda_b,\hat{n}_b)=I$. The direction of the measurement is given by  the unit vector $\hat{n}_b$, which can be parametrized in the spherical polar coordinate as
$\hat n_b = \left(\sin{\theta_b}\cos{\phi_b}, \sin{\theta_b}\sin{\phi_b}, \cos{\theta_b}\right)$, 
where $\forall b$, $\theta_b,\phi_b\in\mathbb{R}$ with   $\theta_b\in[0,\pi]$ and $\phi_b\in[0,2\pi]$} 

\added{We assume the measurements to be affected \emph{only} by white noise on each qubit implemented by a \emph{depolarizing} channel, and write the corresponding POVM element on the assisting qubit $b$ as 
\begin{eqnarray}
    P(\lambda_b,\vec{\eta}_b)&=& \Gamma_{\eta_b}\left(P(\lambda_b,\hat{n}_b)\right), \nonumber\\&=&\eta_b P(\lambda_b,\hat{n}_b)+(1-\eta_b)\frac{I}{2},\nonumber\\ &=&\frac{1}{2}\left[I+\lambda_b\vec{\eta}_b.\vec{\sigma}_b\right],
    \label{eq:povm}
\end{eqnarray}
where $\Gamma_{\eta_b}(.)$ denotes the depolarizing channel of strength $(1-\eta_b)$. Here we 
 define $\vec{\eta}_b=\eta_b\hat{n}_b$ with $\eta_b$ ($0\leq \eta_b\leq 1$) being the \emph{degree of unsharpness} (DoU) of the measurement on the assisting qubit $b$ which quantifies the amount of noise influencing the measurement device. For $\eta_b=1$, one obtains a \emph{sharp} projection measurement along $\hat{n}_b$ on the qubit $b$. The advantage of this formalism lies in the applicability of the results to all environments of the measurement device whose effect on the measurement can be modeled by a depolarizing channel, with the strength of the depolarizing noise quantifying the severity of the effect of the environment on the measurement, without access to microscopic details, or control of the environment.}

Post POVMs on all assisting qubits corresponding to the unsharpness $\eta=\{{\eta}_1,{\eta}_2,\cdots,{\eta}_{N_B}\}$ along the directions $\hat{\mathbf{n}}=\{\hat n_1,\hat n_2,\ldots,\hat n_{N_B}\}$, the $N$-qubit state corresponding to the measurement outcome $\lambda=\{\lambda_1,\lambda_2,\cdots,\lambda_{N_B}\}$ occurring with the probability $p_{(\lambda,\vec\eta)}=\text{Tr}\left[M_{(\lambda,\vec\eta)}\rho M_{(\lambda,\vec\eta)}^\dagger\right]$ reads as
\begin{eqnarray}
    \rho_{(\lambda,\vec\eta)}=p_{(\lambda,\eta)}^{-1}\left[M_{(\lambda,\vec\eta)}\rho M_{(\lambda,\vec\eta)}^\dagger\right],
    \label{eq:post_measured_state_first_round}
\end{eqnarray}
where $M_{(\lambda,\vec\eta)}=\otimes_{b=1}^{N_B}M(\lambda_b,\vec\eta_b)$ with $M(\lambda_b,\vec\eta_b)$ defined by  $P(\lambda_b,\vec{\eta}_b)=M^\dagger(\lambda_b,\vec\eta_b)M(\lambda_b,\vec\eta_b)$,  and $\vec\eta=\{\vec{\eta}_1,\vec{\eta}_2,\cdots,\vec{\eta}_{N_B}\}$. Using the parametrization of $\hat{n}_b$, each POVM element can be represented as 
\begin{eqnarray}
    \nonumber M(\lambda_b, \vec\eta_b)&=&\sqrt{\frac{1+\lambda_b\eta_b}{2}}\ketbra{\chi^+}{\chi^+}_b \nonumber\\ &&+\sqrt{\frac{1-\lambda_b\eta_b}{2}}\ketbra{\chi^-}{\chi^-}_b,
    \label{eq:parametrization_1}
\end{eqnarray}
with
\begin{eqnarray}
    \ket{\chi^{+}}_b&=&\cos{\frac{\theta_b}{2}}\ket{0}_b+e^{i\phi_b}\sin{\frac{\theta_b}{2}}\ket{1}_b, \nonumber\\ 
    \ket{\chi^{-}}_b&=&\sin{\frac{\theta_b}{2}}\ket{0}_b-e^{i\phi_b}\cos{\frac{\theta_b}{2}}\ket{1}_b,
    \label{eq:parametrization_2}
\end{eqnarray}
satisfying the completeness relation $\sum_{\lambda_b}P(\lambda_b, \vec\eta_b)=I$ for a fixed $\vec\eta_b$. Let us denote the ensemble of the post-measured states on $A$ by $\left\{p_{({\lambda,\vec\eta})},\varrho_{(\lambda,\vec\eta)}\right\}$, where the state\footnote{In the rest of the paper, unless otherwise stated, we denote the state of the full system with $\rho$, and the state of the unmeasured subsystem $A$ with $\varrho$.}  $\varrho_{(\lambda,\vec\eta)}=\text{Tr}_B[\rho_{(\lambda,\vec\eta)}]$ on $A$ occurs with probability $p_{({\lambda,\vec\eta})}$.  The maximum average entanglement localized over the subsystem $A$ post this round of measurement is referred to as the \emph{localizable entanglement}, and is given by 
\begin{eqnarray}
    \mathcal{E}_{\vec{\eta}^{\text{ opt}}}(\rho)=\max\overline{E}_{\vec\eta}(\rho),
\end{eqnarray}
with 
\begin{eqnarray}
\overline{E}_{\vec\eta}(\rho)=\sum_{\lambda}p_{(\lambda,\vec\eta)}E\left(\varrho_{(\lambda,\vec\eta)}\right).  
\end{eqnarray}
Here,  the maximization is carried  over all possible POVMs with unsharpness $\eta$ on all the assisting qubits in $B$, and $E$ is a pre-decided entanglement measure -- bipartite or multipartite -- quantifying entanglement in a state on $A$. It is worthwhile to note here that the usual definition of LE~\cite{divincenzo1998, smolin2005, popp2005, gour2006} corresponds to the case where $\eta_b=1$ $\forall b\in B$, i.e., sharp projection measurements are performed on all assisting qubits, such that  $\vec{\eta}_b=\hat{n}_b$ $\forall b\in B$, implying $\vec{\eta}^{\text{ opt}}=\hat{n}^{\text{opt}}$ (see Eq.~(\ref{eq:povm})), and we denote it by $\mathcal{E}_{\hat{n}^{\text{opt}}}(\rho)$, with $\mathcal{E}_{\hat{n}^{\text{opt}}}(\rho)> \mathcal{E}_{\vec{\eta}^{\text{ opt}}}(\rho)$.      


In this paper, we focus on localizing bipartite entanglement over a given bipartition, say, $A_1:A_2$ of the subsystem $A$. Towards this, we employ negativity~\cite{vidal2002} as the chosen entanglement measure $E$, quantifying entanglement using the partial transposition  criteria \cite{Horodecki:1997vt,PhysRevLett.77.1413}. It is defined as $E(\varrho)=\sum_{i}|e_i|$, where $\{e_i\}$ is the set of all negative eigenvalues of the partially transposed state $\varrho^{T_{A_1}}$, the partial transposition being taken with respect to the partition $A_1$ of $A$. In the case of two qubits, non-vanishing negativity over the partition $A_1:A_2$ is a necessary and sufficient condition for establishing entanglement~\cite{Horodecki96}. 

\begin{figure*}
    \centering
    \includegraphics[width=0.7\linewidth]{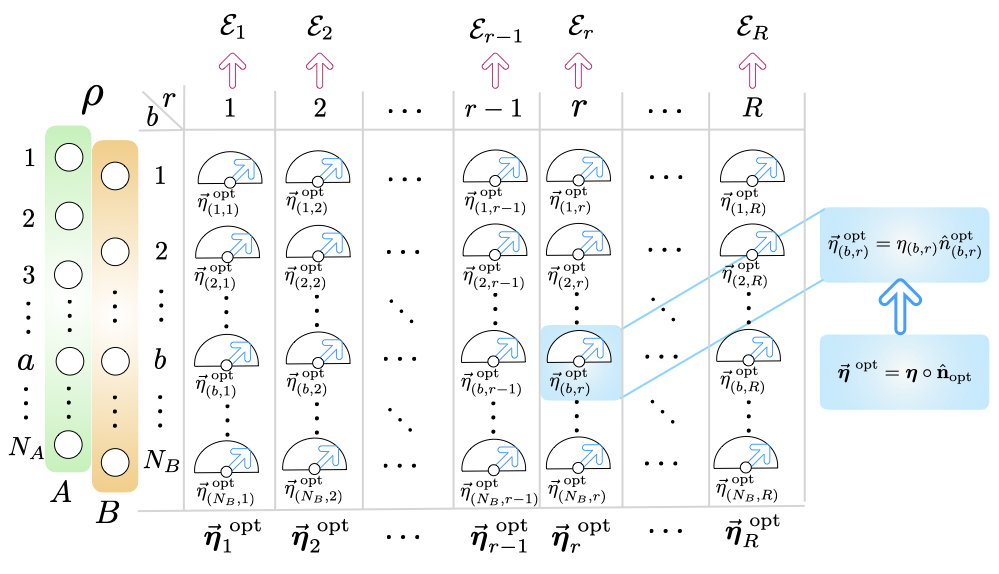}
    \caption{\textbf{Repeated POVMs and sequentially optimized localizable entanglement.} A set of $R$ rounds of POVMs on each of the $N_B$ qubits in the subsystem $B$ of a quantum state $\rho$ is performed. The overall unsharpness corresponding to these measurements can be represented by the $N_B\times R$ UM $\boldsymbol{\eta}$ with elements $\eta_{(b,r)}$, while the MM $\hat{\mathbf{n}}^{\text{opt}}$ with elements $\hat{n}^{\text{opt}}_{(b,r)}$ provides the optimal measurement on all qubits in $B$ for all rounds of measurement. These two matrices constitute $\vec{\boldsymbol{\eta}}^{\text{ opt}}=\boldsymbol{\eta}\circ\hat{\mathbf{n}}_{\text{opt}}$ with elements $\vec{\eta}^{\text{ opt}}_{(b,r)}=\eta_{(b,r)}\hat{n}_{(b,r)}^{\text{opt}}$, whose columns $\vec{\boldsymbol{\eta}}_r^{\text{ opt}}=\boldsymbol{\eta}_r\circ\hat{\mathbf{n}}_{r}$ provides the degree of unsharpness as well as the optimal measurement setup corresponding to the $r$th round of measurement.  After each measurement round $r$, one can compute the corresponding localizable entanglement $\mathcal{E}_{r}^{\text{seq}}$ following the sequentially optimized localization approach discussed in Sec.~\ref{subsec:comment_on_optimization}.}
    \label{fig:POVM}
\end{figure*}

\subsection{Localization via multiple rounds of POVMs} 
\label{subsec:multiple_rounds}

Consider the case of an $N$-qubit system in the state $\rho$ with bipartition $A:B$, where $B$ ($A$) is constituted of $1$ ($N-1$) qubit(s). Operationally, for a specific value of the DoU $\eta_b$, determination of LE over the subsystem $A$ involves (a) creating a large number of identical copies of $\rho$, (b) preparing the complete post-measured ensemble on $A$ via performing all possible single-qubit POVMs on the assisting qubit, and (c) choosing the optimal POVM (i.e., optimal $\hat{n}_{b}^{\text{opt}}$) on $b$ that provides the maximum value of $\overline{E}_{\vec{\eta}}(\rho)$, with $\vec{\eta}\rightarrow\vec{\eta}^{\text{ opt}}=\eta\hat{n}_b^{\text{opt}}$. Note that post a sharp projection measurement in any direction $\hat{n}_b$, entanglement between the partitions $A$ and $B$ in each post-measured state obtained from each copy of $\rho$ vanishes completely, thereby rendering the assisting qubit to be \emph{useless} for any further \emph{assist} in concentrating entanglement over $A$  in any one of the post-measured states. In contrast, single-qubit POVMs on $b$ in any direction may leave residual entanglement between the partitions $A$ and $B$ in the post-measured states, which introduces the possibility of using the assisting qubit $b$ for further cooperation on any chosen post-measured state. Motivated by this, in the case of POVMs, we ask the following question:
\emph{Can the entanglement localizable  via sharp projection measurements on all assisting qubits be achieved using  multiple rounds of unsharp measurements on each of the assisting qubits?}  Here, we assume that the DoU corresponding to a measurement on an assisting qubit is a property of the measurement apparatus, and is fixed during one round of measurement. However, it can change in between two consecutive rounds owing to the possibility of an improvement, or a deterioration of the apparatus.

To systematically investigate the above question, we introduce the index $r$ representing the \emph{rounds of measurement}. For efficient representation of multiple rounds of POVMs on multiple assisting qubits, we define the following:
\begin{enumerate}
    \item[(1)] The representative unsharpness of a sequence of, say, $R$ measurements on all qubits in $B$ can be imagined as an $N_B\times R$ matrix $\boldsymbol{\eta}$, referred to as the \emph{unsharpness matrix} (UM), whose element $\eta_{(b,r)}$ provides the DoUs corresponding to the $r$th measurement on the qubit $b$ (see Fig.~\ref{fig:POVM}). Here, $1\leq b\leq N_B$, and $1\leq r \leq R$. 
    \item[(b)] In a similar fashion, the representative measurement direction of the sequence of $R$ measurements on all qubits in $B$ is given by the $N_B\times R$ matrix $\hat{\mathbf{n}}$, referred to as the \emph{measurement matrix} (MM), whose element $\hat{n}_{(b,r)}$  represents the measurement direction when the qubit $b$ is measured in the $r$th round of measurement. See Fig.~\ref{fig:POVM} for an illustration. 
    \item[(c)] The measurement outcomes obtained after $R$ rounds of measurement can also be imagined in the form of an $N_B\times R$ \emph{outcome matrix} (OM) $\boldsymbol{\lambda}$, where the element $\lambda_{(b,r)}=\pm 1$ represents the outcome of the $r$th measurement on the qubit $b$. After $R$ rounds of measurements, the set of all possible outcomes up to this round is denoted by $\Lambda_R$, hosting the $2^{RN_B}$ possible $\boldsymbol{\lambda}$ matrices. 
\end{enumerate}
Note that one can also define the matrix $\vec{\boldsymbol{\eta}}=\eta\circ\hat{\mathbf{n}}$ with elements $\vec{\eta}_{(b,r)}=\eta_{(b,r)}\hat{n}_{(b,r)}$, and identify the sequence $\vec\eta=\{\vec\eta_1,\vec\eta_2,\cdots,\vec\eta_{N_B}\}$  considered in the case of single-round measurements (see Sec.~\ref{subsec:single_round}) as the $N_B\times 1$ column matrix $\vec{\boldsymbol{\eta}}$. 
 
In the above notations, the post-measured $N$-qubit state after $R$ rounds of measurements on each of $N_B$ assisting qubits is given by 
\begin{eqnarray}
    \rho_{(\boldsymbol{\lambda},\vec{\boldsymbol{\eta}})}=p_{(\boldsymbol{\lambda},\vec{\boldsymbol{\eta}})}^{-1}\left[M_{(\boldsymbol{\lambda},\vec{\boldsymbol{\eta}})}\rho M_{(\boldsymbol{\lambda},\vec{\boldsymbol{\eta}})}^\dagger\right],
    \label{eq:post_measured_state_R_round}
\end{eqnarray}
occurring with the probability 
\begin{eqnarray}
    p_{(\boldsymbol{\lambda},\vec{\boldsymbol{\eta}})}=\text{Tr}\left[M_{(\boldsymbol{\lambda},\vec{\boldsymbol{\eta}})}\rho M_{(\boldsymbol{\lambda},\vec{\boldsymbol{\eta}})}^\dagger\right],
\end{eqnarray}
where 
\begin{eqnarray}
    M_{(\boldsymbol{\lambda},\vec{\boldsymbol{\eta}})}=\bigotimes_{b=1}^{N_B}\bigotimes_{r=1}^{R}M(\lambda_{(b,r)},\vec\eta_{(b,r)})
\end{eqnarray} 
with 
\begin{eqnarray}
    P(\lambda_{(b,r)},\vec{\eta}_{(b,r)})=M^\dagger(\lambda_{(b,r)},\vec\eta_{(b,r)})M(\lambda_{b,r},\vec\eta_{(b,r)}).
\end{eqnarray} 
For a fixed $\boldsymbol{\eta}$,  the LE, after $R$ rounds, can then be defined as  
\begin{eqnarray}
    \mathcal{E}_{R}(\rho)=\max_{\mathcal{P}}\overline E_{R}(\rho_{(\boldsymbol{\lambda},\vec{\boldsymbol{\eta}})}),
    \label{eq:le_overall}
\end{eqnarray}
with  
\begin{eqnarray}
    \overline{E}_{R}(\rho_{(\boldsymbol{\lambda},\vec{\boldsymbol{\eta}})})=\sum_{\Lambda_R}p_{(\boldsymbol{\lambda},\vec{\boldsymbol{\eta}})}E(\varrho_{(\boldsymbol{\lambda},\vec{\boldsymbol{\eta}})}), 
    \label{eq:av_ent_overall}
\end{eqnarray}
where $\varrho_{(\boldsymbol{\lambda},\vec{\boldsymbol{\eta}})}=\text{Tr}_{B}[\rho_{(\boldsymbol{\lambda},\vec{\boldsymbol{\eta}})}]$ is the post-measured state on $A$ corresponding to the OM $\boldsymbol{\lambda}$ occurring with the probability $p_{(\boldsymbol{\lambda},\vec{\boldsymbol{\eta}})}$. In Eq.~(\ref{eq:le_overall}), the maximization is performed over the complete set $\mathcal{P}$ of all possible POVM directions, $\hat{\mathbf{n}}$. Note that the optimization for $R$ rounds of measurement on each of the $N_B$ qubits is performed \emph{simultaneously} here, which corresponds to a $2RN_B$-parameter maximization problem. We refer to the LE maximized in this fashion as the \emph{globally optimized} LE (GLE), and denote the optimum MM leading to the maximum value of $\mathcal{E}_R$ by $\hat{\mathbf{n}}^{\text{opt}}$. Note further that $\mathcal{E}_R$ is  indeed a function of the DoUs corresponding to the POVMs in the $R$ rounds of measurements on all $N_B$ qubits. However, for brevity, we refrain from including the degree of unsharpness in the notation for LE.

Let us now consider a given $N$-qubit state $\rho$ for which the entanglement localizable over the subsystem $A$ via \emph{a single round of sharp single-qubit projection measurements} on all assisting qubits is $\mathcal{E}_{\hat{m}^{\text{opt}}}(\rho)$ (see Sec.~\ref{subsec:single_round}), where $\hat{m}^{\text{opt}}$ is the optimal measurement directions on the assisting qubits. With the definition of LE using multiple rounds of unsharp measurements and sequential optimization presented above, we rephrase the question posed at the beginning of Sec.~\ref{subsec:multiple_rounds} as the following.
\begin{enumerate}
    \item \emph{Since noisy measurement is inevitable, can one 
    achieve the ideal case, i.e., } $\mathcal{E}_{R}=\mathcal{E}_{\hat{m}^{\text{opt}}}$ \emph{with $R$ rounds of noisy measurements on all assisting qubits?} 
    \item \emph{Given an affirmative answer to the previous question, is it possible to recognize a pattern in the optimal measurements of different rounds of unsharp measurements?}  
\end{enumerate}
In the subsequent sections, we investigate these questions in detail for a number of paradigmatic class of states shared in a network.

\section{Advantages from multiple rounds of measurements in localizing entanglement}
\label{sec:three_qubit}

In this section, we demonstrate that while single round of noisy measurement has detrimental effect on localizing entanglement, multiple rounds of noisy measurements can facilitate achieving $\mathcal{E}_{R}=\mathcal{E}_{\hat{m}^{\text{opt}}}$. As mentioned in the introduction, concentration of entanglement via sequential noisy measurements may provide higher controlling power for the assisting qubits than a scheme with a single PV measurement. We also probe the patterns in optimal measurements in multiple rounds of measurements, which is advantageous in deciding the measurement direction in the next round after a round of measurement is performed. We also comment on the use of the multiple rounds of noisy measurements in preparing highly entangled two-qubit quantum states starting from three-qubit GHZ and W states.

\subsection{Sequential maximization vs maximization over all rounds}
\label{subsec:comment_on_optimization}

Let us now discuss how the optimization is performed for a fixed state. First, notice that the optimization involved in the definition of LE is a $2RN_B$ parameter maximization problem. While numerically performing the maximization is feasible for moderate values of $R$ and $N_B$, analytical maximization of LE over all $2RN_B$ real parameters in its full generality for a given $\boldsymbol{\eta}$ matrix is non-trivial even for small values of $R$ and $N_B$. To overcome this difficulty, we consider a round-wise maximization of the LE, where maximization is performed separately for, say, the second round of measurements on $N_B$ assisting qubits, starting from all of the $2^{N_B}$ post-measured states corresponding to the \emph{optimal measurement} in the first round, and so on. Borrowing the notations introduced  in Sec.~\ref{subsec:multiple_rounds}, note that the post-measured $N$-qubit states $\{\rho_r\}$ after the $r$th round of measurements, starting from all the post-measured $N$-qubit states $\{\rho_{r-1}\}$ corresponding to the optimal measurement $\hat{\mathbf{u}}_r$ of the $(r-1)$th round performed over $N_B$ qubits, is given by 
\begin{eqnarray}
    \rho_{r}&=&p_{r}^{-1}\left[M_{(\boldsymbol{\lambda}_{r},\vec{\boldsymbol{\eta}}_r)}\rho_{r-1}M^\dagger_{(\boldsymbol{\lambda}_{r},\vec{\boldsymbol{\eta}}_r)}\right],
\end{eqnarray} 
occurring with the probability 
\begin{eqnarray}
    p_{r} &=&\text{Tr}\left[M_{(\boldsymbol{\lambda}_{r},\vec{\boldsymbol{\eta}}_r)}\rho_{r-1}M^\dagger_{(\boldsymbol{\lambda}_{r},\vec{\boldsymbol{\eta}}_r)}\right]. 
\end{eqnarray}
Here,  $M_{(\boldsymbol{\lambda}_r,\vec{\boldsymbol{\eta}}_r)}=\otimes_{b=1}^{N_B}M(\lambda_{(b,r)},\vec\eta_{(b,r)})$ is defined corresponding to the $r$th columns of the $\vec{\boldsymbol{\eta}}$ and the $\boldsymbol{\lambda}$ matrices, where $\vec{\boldsymbol{\eta}}=\boldsymbol{\eta}\circ\hat{\mathbf{u}}$ with elements $\vec{\eta}_{(b,r)}=\eta_{(b,r)}\hat{u}_{(b,r)}$, with $\hat{u}_{(b,r)}$ representing the optimal measurement direction corresponding to the $r$th round of POVM on the qubit $b$. Note that in this approach, the optimal MM $\hat{\mathbf{u}}$ is not known beforehand, and is constructed by appending the columns $\hat{\mathbf{u}}_r$ corresponding to the optimal measurement on all qubits in $B$ in the round $r$. The LE on $A$ in the $r$th round of measurements on the qubits in $B$ can be represented as
\begin{eqnarray}
    \mathcal{E}_{r}^{\text{seq}}=\max\overline E_{r},
    \label{eq:le_r}
\end{eqnarray}
with 
\begin{eqnarray}
    \overline{E}_{r}=\sum_{\{\boldsymbol{\lambda}_r\}}p_{r}E(\varrho_{r}),
    \label{eq:av_ent_r}
\end{eqnarray}
where $\varrho_r=\text{Tr}_B(\rho_r)$ is the post-measured state on $A$, and $\{\boldsymbol{\lambda}_r\}$ is the set of $r$th columns of all $\boldsymbol{\lambda}\in \Lambda_R$. 
Here, the maximization is carried out over all possible POVMs in the $r$th round of measurements over the assisting qubits in $B$, which reduces to a $2N_B$-parameters optimization problem (see Sec.~\ref{subsec:single_round}) in each round. Given an $N$-qubit initial state $\rho$, in this approch, one can, in principle, determine  $\mathcal{E}_{r}^{\text{seq}}$ after each round $r$ of measurements (see Fig.~\ref{fig:POVM}). We denote the LE obtained after $R$ rounds of such measurements and sequential optimization by $\mathcal{E}_R^{\text{seq}}$, and refer it as the \emph{sequentially-optimized} LE (SLE).

While analytically tackling SLE is less cumbersome compared to the GLE, due to the reduced number of parameters involved in maximization in each round of measurement, it is not at all clear (a) whether $\mathcal{E}_{R}^{\text{seq}}$ and $\mathcal{E}_R$ are equal, and (b)  if $\hat{\mathbf{u}}=\hat{\mathbf{n}}^{\text{opt}}$. However, our numerical investigation for paradigmatic multiqubit pure states reveal affirmative answer to both of these questions, thereby providing an avenue to determine closed forms of LE involving the DoUs and the state parameters.

\subsection{Effectiveness of noisy measurements with multiple rounds: Generalized GHZ and W states}
\label{subsec:gGHZ}

We now investigate the LE computed via multiple rounds of noisy measurements on a qubit in three-qubit gGHZ and gW states. Unless otherwise stated, we always use qubit $b=3$ for assistance, and localize bipartite entanglement as quantified by negativity over the qubits $a=1$ and $a=2$. 

\subsubsection{Generalized GHZ states}
We start with the three-qubit gGHZ states \cite{Greenberger2007} given by 
\begin{eqnarray}
    \ket{\text{gGHZ}}&=& c_{0}\ket{000}+c_{1}\ket{111},
    \label{eq:ghz_state_ex}
\end{eqnarray}
with $c_{0},c_{1}\in \mathbb{C}$, and $|c_{0}|^{2}+|c_{1}|^{2}=1$. For sharp projective measurements on qubit $3$ $(\eta_{3}=1)$, $\hat{m}^{\text{opt}}=\{\hat{m}_3^{\text{opt}}\}$ is independent of $\phi_3$, and lies on the great circle of the Bloch sphere, characterized only by $\theta_3=\pi/2$ (see Sec.~\ref{subsec:multiple_rounds}), with $\mathcal{E}_{\hat{m}_3^{\text{opt}}}=|c_0c_1|$~\cite{divincenzo1998,Krishnan2023}. In order to investigate the case of $R$ rounds of noisy measurements on qubit $3$, for simplifying the calculation, we make the reasonable assumption that the quality of the measurement device does not change between different rounds of measurements, leading to $\eta_{(3,r)}=\eta\;\forall\; r=1,2,\cdots,R$. Note that, even without this assumption, the qualitative results presented here remain unaltered, although, depending on the orderings of $\eta_1,\eta_2,\ldots,$ more additional conditions may appear. Further, we define the difference between $\mathcal{E}^{\text{seq}}_r$ after the $r$th round of measurements on qubit $3$, and $\mathcal{E}_{\hat{m}_3^{\text{opt}}}$, relative to $|c_0c_1|$ as  
\begin{eqnarray}
\Delta_{r}=\frac{\left|\mathcal{E}_{\hat{m}_3^{\text{opt}}}-\mathcal{E}^{\text{seq}}_r\right|}{|c_0c_1|}.
\end{eqnarray}
Using this, we propose the following for multiple rounds of single-qubit noisy measurements on the gGHZ state.
\begin{proposition}\label{prop:gghz}
     For multiple rounds of noisy measurements on a qubit in a three qubit gGHZ state given in Eq.~(\ref{eq:ghz_state_ex}), $\Delta_r\leq \Delta_{r-1}$ $\forall$ $r=1,2,\cdots,R$. 
\end{proposition}
\noindent We now discuss the analysis leading to the  Proposition~\ref{prop:gghz}. Note that post first-round of measurement on the assisting qubit $3$,  one obtains $\overline E_{1}=\eta|c_0c_1||\sin{\theta_{(3,1)}}|$ (see Eq.~(\ref{eq:av_ent_overall})) to be independent of $\phi_{(3,1)}$, where Eqs.~(\ref{eq:parametrization_1}) and (\ref{eq:parametrization_2}) are used. Here, in order to distinguish between the parameters involved in different rounds of measurements, we have adopted notations similar to $\hat{n}_{(b,r)}$ for the real parameters $(\theta,\phi)$ also. Maximization with respect to $\theta_{(3,1)}$ results in $\mathcal{E}_{1}^{\text{seq}}=\eta|c_0c_1|$,  with the optimal $\theta_{(3,1)}=\pi/2$, implying that $\hat{u}_{(3,1)}$ lies along the great circle of the Bloch sphere, which leads to 
\begin{eqnarray}
\Delta_1=1-\eta.
\end{eqnarray}

\begin{figure}
    \centering
    \includegraphics[width=\linewidth]{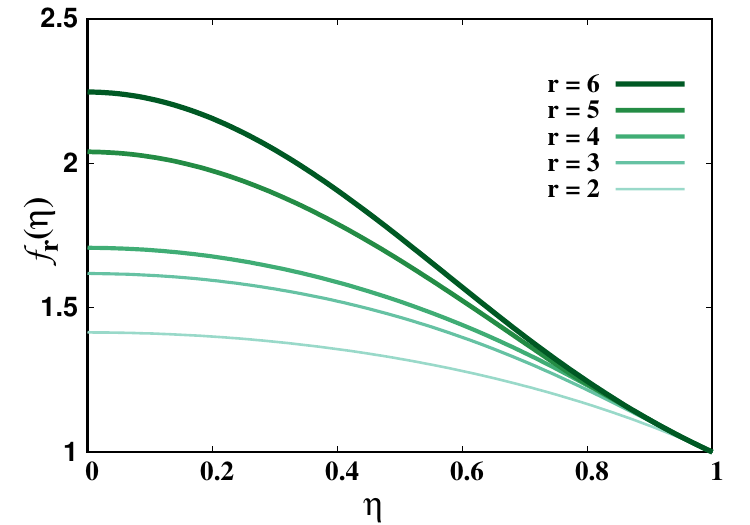}
    \caption{$f_r(\eta)$ in Eq.~\eqref{eq:delta_r} with $\eta$ for different values of $r$ in the range $2\leq r\leq 6$ when the shared state is the $\ket{gGHZ}$ state. Both the axes are dimensionless.}
    \label{fig:eta_function}
\end{figure}

Next, starting from the set of post-measured states corresponding to the optimal measurement $\hat{u}_{(3,1)}$ on the qubit $3$ in the first round, the average entanglement in the second round is written as 
\begin{eqnarray}
    \overline E_{2}=\frac{1}{2}|c_{0}c_{1}|\left(\sqrt{d_{0}}+\sqrt{d_{1}}\right)
\end{eqnarray}
where
\begin{eqnarray}
    \nonumber d_{\alpha}&=&\eta \sin \theta_{(3,2)}[\eta \sin \theta_{(3,2)} (1-\eta^{2} \sin^2\delta\phi_{12})\\ &-&(-1)^{\alpha}2 \eta \cos\delta\phi_{12}]+\eta^{2},
\end{eqnarray}
with $\alpha=0,1$, and  $\delta\phi_{12}=\phi_{(3,1)} -\phi_{(3,2)}$, where we have assumed the UM corresponding to two consecutive rounds of unsharp measurements on the qubit $3$ as $\boldsymbol{\eta}=\begin{bmatrix}\eta & \eta\end{bmatrix}$. Using $\partial\overline{E}_{2}/\partial\theta_{(3,2)}=0$ and $\partial\overline{E}_{2}/\partial\delta\phi_{12}=0$, one obtains $\theta_{(3,2)}=\pi/2$, and $\delta\phi_{12}=\pm\pi/2$. Further, at $(\theta_{(3,2)}=\pi/2,\delta\phi_{12}=\pm\pi/2)$, $\partial^2\overline{E}_2/\partial\theta^{2}_{(3,2)}<0$ and $[\partial^2\overline{E}_2/\partial\theta^{2}_{(3,2)}][\partial^2\overline{E}_2/\partial\delta\phi^{2}_{12}]-[\partial^2\overline{E}_2/\partial\theta_{(3,2)}\partial\delta\phi_{12}]>0$, ensuring $\hat{u}_{(3,2)}=(\theta_{(3,2)}=\pi/2,\phi_{(3,2)}=\phi_{(3,1)}\pm\pi/2)$ to be the optimal direction, with $\hat{u}_{(3,2)}\perp\hat{u}_{(3,1)}$. By substituting the optimal values of $\theta_{(3,2)}$ and $\phi_{(3,2)}$, $\mathcal{E}_{2}^{\text{seq}}$ is obtained as $\mathcal{E}_{2}^{\text{seq}} = \eta\sqrt{2-\eta^{2}}|c_{0}c_{1}|$, with an optimal MM $\hat{\mathbf{u}}=\begin{bmatrix}\hat{u}_{(3,1)} & \hat{u}_{(3,2)}\end{bmatrix}$, where  $\hat{u}_{(3,2)}$ is given by $(\theta_{(3,2)}=\pi/2$, $\phi_{(3,2)}=\phi_{(3,1)}\pm\pi/2)$. Therefore, 
\begin{eqnarray}
\Delta_2= 1-\eta\sqrt{2-\eta^2}. 
\end{eqnarray}
and  $\Delta_1-\Delta_2=\eta\left[\sqrt{2-\eta^2}-1\right]>0$ $\forall$ $\eta$ in the range $0\leq \eta\leq 1$.

\begin{figure}
\includegraphics [width=\linewidth]{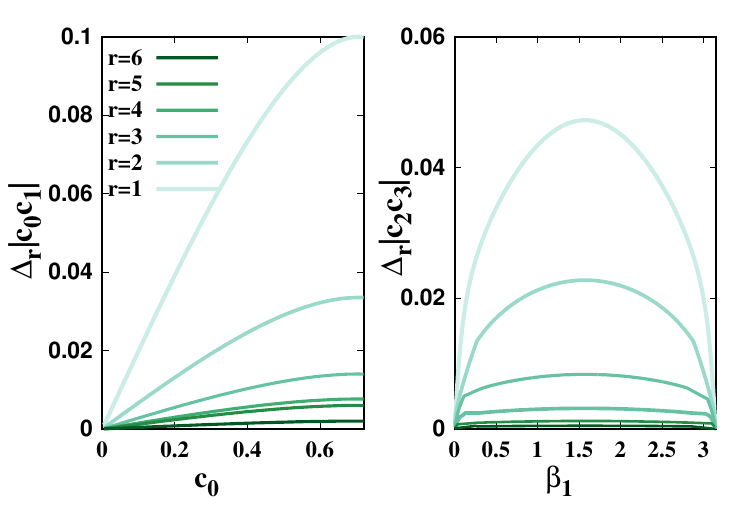}
\caption{Variations of (a) $\Delta_r|c_{0}c_{1}|$ as  a function of $c_0$ in the case of arbitrary gGHZ states (Eq.~(\ref{eq:ghz_state_ex})), and (b) $\Delta_r|c_{2}c_{3}|$  as a function of $\beta_1$ with $\beta_2=\pi/2$ in the case of arbitrary gW states (Eq.~(\ref{eq:gw})),  when $\eta=0.8$, and $1\leq r\leq 6$. All quantities plotted are dimensionless.} 
\label{fig:delta_function}
\end{figure}

Proceeding in a similar fashion, we observe that $\hat{u}_{(3,r)}\perp\hat{u}_{(3,r-1)}$ $\forall$ $r$, and   $\Delta_r$ $(2\leq r\leq R)$ takes the form 
\begin{eqnarray}
    \Delta_{r}=1-\eta f_r(\eta)\;\forall\; r\geq 2,
    \label{eq:delta_r}
\end{eqnarray}
where $f_r(\eta)$ is an analytic monotonically decreasing function of $\eta$ having value $\geq 1$ in the range $0\leq \eta\leq 1$, with the minimum value $1$ occurring at $\eta=1$ $\forall$ $r>1$, and  $f_r(\eta)\geq f_{r-1}(\eta)$ $\forall$ $\eta$ ($0\leq \eta\leq 1$). In Fig.~\ref{fig:eta_function}, we plot $f_r(\eta)$ for $2\leq r\leq 6$ for demonstration, although we do not write the explicit forms of $f_r(\eta)$ to keep the text uncluttered. \added{Note that the case $\eta = 0$ corresponds to maximally noisy measurements on the assisting qubit(s), implying $\mathcal{E}_r^{\text{seq}}=0$, and thereby $\Delta_r=1$ $\forall r$, which is ensured in Eq.~(\ref{eq:delta_r}) by $\eta=0$ irrespective of the values of $f_r(\eta)$ at $\eta=0$}. In Fig.~\ref{fig:delta_function}(a), we plot the variations of $\Delta_r|c_0 c_1|$ for $1\leq r\leq R=6$  as a function of $c_0$ with $\eta$ fixed at $\eta=0.8$. It is clear from the figure that for all values of $\eta$, $\Delta_r\rightarrow 0$ as $r$ increases, and hence $\mathcal{E}_r^{\text{seq}}\rightarrow\mathcal{E}_{\hat{m}_3^{\text{opt}}}$, which leads to the following observation for the three-qubit gGHZ states.

\noindent\textbf{Observation 1.} \emph{For a three-qubit gGHZ state, the LE corresponding to (noiseless) projection measurements on the assisting qubit $3$ can be attained by at most six rounds of noisy measurements on the qubit $3$ up to an error $\mathcal{E}_{\hat{m}_3^{\text{opt}}}-\mathcal{E}_{6}^{\text{seq}}\lesssim 5\times 10^{-3}$.}

\noindent Note here that due to the symmetry of the gGHZ, if measurement is performed on other qubits instead of qubit $3$, the result remains unaltered.

\subsubsection{Generalized W states}

We consider another family of three-qubit states, namely gW states, given by
\begin{eqnarray}
\ket{\text{gW}}&=&c_1\ket{001}+c_2\ket{010}+c_3\ket{100},
\label{eq:gw}
\end{eqnarray}
where $c_i\in\mathbb{C}$ $\forall$ $i=1,2,3$, and $\sum_{i=1}^3|c_i|^2=1$ and compare its behavior with gGHZ. In this case, noiseless projection measurement along any arbitrary direction $\hat{m}$ on the assisting qubit $3$ provides the LE over qubits $1$ and $2$ as
$\mathcal{E}_{\hat{m}_3^{\text{opt}}}=|c_2c_3|$ \cite{Krishnan2023}, when negativity is used as entanglement measure over the qubits $1$ and $2$. On the other hand, for multiple rounds of noisy measurements on qubit $3$ with the same DoU $\eta$ for each round of measurements, similar to the three-qubit gGHZ state, one can define
\begin{eqnarray}
    \Delta_r&=&\frac{\left|\mathcal{E}_{\hat{m}_3^{\text{opt}}}-\mathcal{E}^{\text{seq}}_r\right|}{|c_2c_3|}, 
\end{eqnarray}
which leads to the following. 
\begin{proposition}\label{prop:gw}
     For multiple rounds of noisy measurements on a qubit in a three qubit gW state given in Eq.~(\ref{eq:gw}), $\Delta_r\leq \Delta_{r-1}$ $\forall$ $r=1,2,\cdots,R$. 
\end{proposition}
\noindent The analysis of $\Delta_r$ for the three-qubit gW states is similar to the case of the three-qubit gGHZ states. Parametrizing the three-qubit gW states as $c_1=\cos\frac{\beta_2}{2}$, $c_2=\sin\frac{\beta_2}{2}\cos\frac{\beta_1}{2}$, and $c_3=\sin\frac{\beta_2}{2}\sin\frac{\beta_1}{2}$,  we plot $\Delta_r|c_2 c_3|$ as a function of $\beta_1$ for fixed values of $\beta_2=\pi/2$ in Fig.~\ref{fig:delta_function}(b) in support of Proposition~\ref{prop:gw}. The trends of $\delta_r$ remains same for all values of $(\beta_1,\beta_2)$ pair. Also, similar to the three-qubit gGHZ states,  the following observation from Fig.~\ref{fig:delta_function}(b) can be safely made. 

\noindent\textbf{Observation 2.} \emph{For a three-qubit gW state, the LE corresponding to noiseless projection measurements on the assisting qubit $3$ can be attained by at most four rounds of noisy measurements on the qubit $3$ up to an error $\mathcal{E}_{\hat{m}_3^{\text{opt}}}-\mathcal{E}_{4}^{\text{seq}}\lesssim 5 \times 10^{-3}$.}

\begin{figure}
\includegraphics [width=\linewidth]{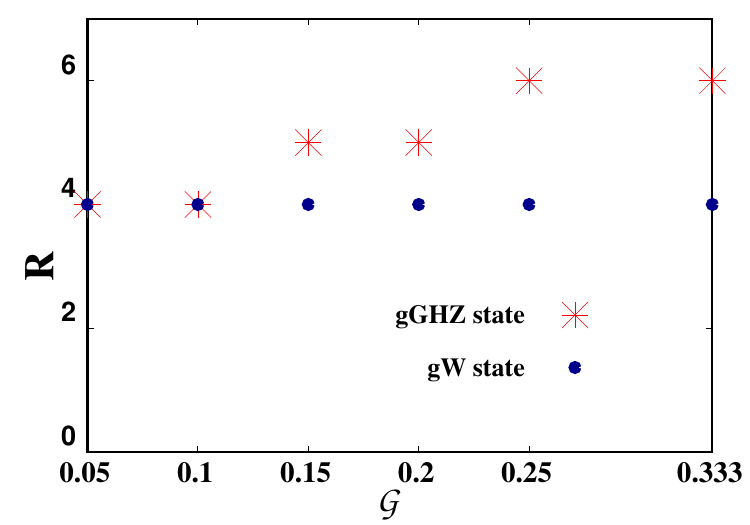}
\caption{The number of measurement rounds, $r$ with respect to the fixed initial entanglement content, \(\mathcal{G}\), measured by GGM, obtained from  gGHZ (squares) and gW (circles) states. In this representation, the blue dotted line indicates how various gW states reach \(\mathcal{E}_{\hat{m}_3^{\text{opt}}}\) , while the green dotted line corresponds to the same for different gGHZ states. Both axis are dimensionless.} 
\label{fig:gghz_gw_initial_ent}
\end{figure}

Notice that for a fixed error, gW states requires less number of measurement rounds than that corresponding to the gGHZ states for achieving $\mathcal{E}_{\hat{m}^{\text{opt}}}$. Since the entanglement properties of the gGHZ and gW states are distinctly different,  we now ask whether the difference in the number of measurement rounds is connected to the multipartite entanglement content of the given state. To address this question, we choose the state parameters $(c_0,c_1)$ of the gGHZ state, and $(c_1,c_2,c_3)$ of the gW state in such a way that the genuinely multipartite entanglement content measured by generalized geometric measure (GGM), $\mathcal{G}$ \cite{PhysRevA.81.012308,PhysRevA.90.032301,PhysRevA.94.022336}, is same, i.e., $\mathcal{G}(\ket{\text{gGHZ}})_{(c_0,c_1)}=\mathcal{G}(\ket{\text{gW}})_{(c_1,c_2,c_3)}$. We observe that the for such a pair of a gGHZ and gW states, the number of rounds of noisy measurements required to achieve $\mathcal{E}_{\hat{m}_3^{\text{opt}}}$ up to an error of $5\times 10^{-3}$ are different, as clearly demonstrated in Fig.~\ref{fig:gghz_gw_initial_ent}, indicating a more intricate mechanism determining the required number of rounds of measurements in play.

In the subsequent sections, we discuss the generality of the Observations 1 and 2 in the case of three-qubit arbitrary pure states, and multiqubit states of systems having four, or higher number of qubits.

\subsubsection{Patterns in optimal measurements}

The set of optimal POVM directions for the first round of noisy measurement on a three-qubit arbitrary gGHZ state is an infinite set, spanning all possible directions along the $x-y$ plane of the Bloch sphere, which, in turn, makes the set of optimal POVM directions for all rounds of measurements an infinite set. However, from the analysis around the first and second round of measurements on a qubit in the gGHZ states, we point out that choosing $\phi_{(3,1)}=0$ and $\pi/2$ fixes the  corresponding $1\times R$ ($R=6$) optimum MM $\hat{\mathbf{u}}$ to be
\begin{eqnarray}
    \hat{\mathbf{u}}=\left\{
    \begin{array}{cc}
     \begin{bmatrix}
        \hat{x} & \hat{y} & \hat{x} & \hat{y} & \hat{x} & \hat{y} 
    \end{bmatrix}    & \text{ for } \phi_{(3,1)}=0 \\
    & \\
    \begin{bmatrix}
        \hat{y} & \hat{x} & \hat{y} & \hat{x} & \hat{y} & \hat{x} 
    \end{bmatrix}     & \text{ for } \phi_{(3,1)}=\pi/2 
    \end{array}\right.
    \label{eq:optimal_mm_pattern_gghz}
\end{eqnarray}
irrespective of the values of $\eta$.   Here, the measurement directions $\hat{x}$ and $\hat{y}$ for the $r$th measurement on qubit $b$  correspond respectively to $\sigma^x$ and $\sigma^y$ measurements, parameterized by
\begin{eqnarray}
    \sigma^x &:& \theta_{(b,r)}=\frac{\pi}{2},\phi_{(b,r)}=0,
    \label{eq:xmeasurement}
\end{eqnarray}
and
\begin{eqnarray}
    \sigma^y &:& \theta_{(b,r)}=\frac{\pi}{2},\phi_{(b,r)}=\frac{\pi}{2},
    \label{eq:ymeasurement}
\end{eqnarray}
respectively.

Determining the optimal MM $\hat{\mathbf{u}}$ in the case of gW states in its full generality under multiple rounds of noisy measurements and for arbitrary DoU $\eta$ is a non-trivial task due to the dependence of the optimal POVM direction on the value of $\eta$. However, for the three-qubit W state given by $c_i=1/\sqrt{3}$ $\forall$ $i=1,2,3$ in Eq.~(\ref{eq:gw}), for arbitrary $\eta$, we find \emph{two of the optimum} MM $\hat{\mathbf{u}}$ to be 
\begin{eqnarray}
    \hat{\mathbf{u}}&=& \begin{bmatrix}
        \hat{z} & \hat{x} & \hat{y} & \hat{z}
    \end{bmatrix}
    \label{eq:optimal_mm_pattern_w_1}
\end{eqnarray}
and 
\begin{eqnarray}
    \hat{\mathbf{u}}&=& \begin{bmatrix}
        \hat{z} & \hat{x} & \hat{z} & \hat{x} 
    \end{bmatrix}
    \label{eq:optimal_mm_pattern_w_2}
\end{eqnarray}
for $R=4$, where $\hat{x}$ and $\hat{y}$ correspond to the $\sigma^x$ and $\sigma^y$ measurements respectively (see Eqs.~(\ref{eq:xmeasurement})-(\ref{eq:ymeasurement})), and $\hat{z}$ stands for a $\sigma^z$ measurement, parameterized by 
\begin{eqnarray}
    \sigma^z &:& \theta_{(b,r)}=\phi_{(b,r)}=0.
    \label{eq:zmeasurement}
\end{eqnarray}

Motivated by the above observations, we construct the set $\mathcal{S}_\perp$ of all possible $1\times 4$ MMs $\hat{\mathbf{u}}$ such that 
\begin{eqnarray}
    \hat{u}_{(3,r)}&\in&\{\hat{x},\hat{y},\hat{z}\},
    \hat{u}_{(3,r-1)}\perp\hat{u}_{(3,r)}\perp\hat{u}_{(3,r+1)},
    \label{eq:pattern_single_qubit}
\end{eqnarray}
and refer to it as the \emph{orthogonal Pauli set} (OPS). Our numerical analysis finds, for each gW state in a sample of $10^4$ Haar-uniformly generated \cite{Bengtsson2006} gW states of the form (\ref{eq:gw}), that an optimization of the LE  over $R=4$ consecutive rounds of noisy measurements with the MM drawn from the set $\mathcal{S}_\perp$ provides the same value of $\mathcal{E}_4^{\text{seq}}$ as in the case where no restriction over the MM is imposed. While the recognition of this pattern of optimal MM considerably reduces the computational effort required for calculating $\mathcal{E}_R^{\text{seq}}$, the question regarding its relevance for arbitrary pure states remains, which we address in the subsequent subsections.

\subsection{More adverse impacts of noisy measurements on   GHZ class than W class}
\label{subsec:ghz_class}

We now investigate whether the features of multiple rounds of noisy measurements observed for the three-qubit gGHZ and gW states are also present for the states chosen from the GHZ and the W classes.  The three-qubit GHZ class of states are given by \cite{Yang2009}, $\ket{\Psi}=\sum_{i=0}^{7}c_{i}\ket{\psi_i}$,
where $c_{i}\in \mathbb{C}$ $\forall$ $i=0,1,\cdots,7$, $\sum_{i=0}^7|c_i|^2=1$, and $\{\ket{\psi_i}\}$ represent  the standard computational basis in the Hilbert space of three qubits. On the other hand, the  three-qubit W class states can be represented as \cite{Yang2009} $\ket{\Phi}=c_0\ket{000}+c_1\ket{100}+c_2\ket{010}+c_3\ket{001}$, where $c_{i}\in \mathbb{C}$ $\forall$ $i=0,1,\cdots,3$, and $\sum_{i=0}^3|c_i|^2=1$. The   GHZ and the W classes of states are shown to be mutually disjoint under stochastic local operations and classical communication~\cite{Durvidal2002}.   

\begin{figure}
\includegraphics [width=\linewidth]{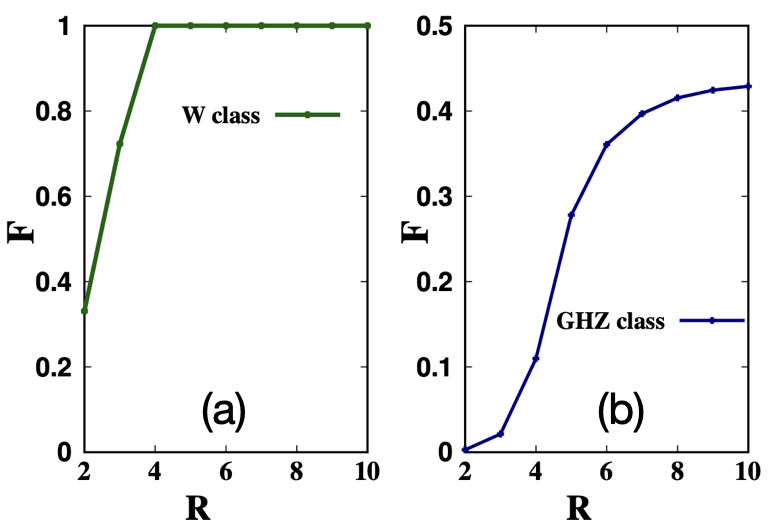}
\caption{\added{Normalized fraction of states, $F$ such that $R$ in the case of three-qubit pure states belonging to  the W class (a) and the GHZ class (b).} In this error window, all states simulated $(5\times 10^4)$ from the W class achieves $\mathcal{E}_{\hat{m}_3^{\text{opt}}}$ after at most $4$ rounds while even after $10$ rounds, only \(42.89 \%\) states from the GHZ class attains the value with sharp measurements. Here, $\eta=0.8$ is taken for all the rounds. All quantities plotted are dimensionless.} 
\label{fig:fraction_R}
\end{figure}

To test whether $\mathcal{E}_R^{\text{seq}}=\mathcal{E}_{\hat{m}_3^{\text{opt}}}$ with a given value of $R$ for an arbitrary three-qubit pure state, we restrict ourselves to noisy Pauli measurements only. For a fixed $R$, we compute the normalized fraction $F$ of states belonging to each class for which $\mathcal{E}_{\hat{m}_3^{\text{opt}}}-\mathcal{E}_{R}^{\text{seq}}\leq 5\times 10^{-3}$, up to our numerical accuracy, where a Haar uniformly generated sample of $5\times 10^4$ states from each of the GHZ and the W class is used and the normalization is performed by dividing over the total number of states simulated. In Fig.~\ref{fig:fraction_R}, we plot $F$ as a function of $R$. In the case of states from the W class, $F$ increases monotonically with increasing $R$, and reaches $F=1$ (up to $5\times 10^{-3}$) at $R\geq 4$, similar to the three-qubit gGHZ and gW states. In particular, in the case of states from the GHZ class, $F$ increases slowly with $R$ such that $\mathcal{E}_R^{\text{seq}}$ of only around $42.\%$ of the GHZ class states approaches $\mathcal{E}_{\hat{m}^{\text{opt}}}$ within a window of $5\times 10^{-3}$ with \(R=10\). We further test the suitability of orthogonal Pauli set to be used for saturating the LE over $4$ rounds, and find it to do this successfully in the case of states belonging to the W class. In contrast, for states from the GHZ class, such patterns of optimal POVM direction is absent. \added{These results remain invariant irrespective of the qubit on which the measurement is performed, indicating that a choice of random pair of qubits for localizing the same entanglement as that obtained by a sharp optimal measurement keeps the above results unaltered (cf.~\cite{Fortescue2007})}. The beneficial role of W-class states with respect to sequential noisy measurements indicate that in networks, sharing states close to W class states are more appropriate than that of the states from the GHZ class. This is due to the fact that more rounds of measurements to reach the optimal value give rise to the possibility of more noise entering the system, thereby affecting the performance of quantum information protocols.

\subsection{Preparing maximally entangled states using sequential measurements}
\label{sec:preparation}
In the process of localizing maximum entanglement over any two qubits in a three-qubit GHZ state via sharp projection measurement along $\hat{x}$ on an assisting qubit, one creates a post-measured ensemble of two maximally entangled states $(\ket{00}\pm\ket{11})/\sqrt{2}$ on the other two qubits, where each state occurs with a probability $1/2$. On the other hand, in the case of the three-qubit W state, maximum LE over any two qubits corresponds to sharp projection measurement in arbitrary direction as long as negativity is used as entanglement measure. Choosing the measurement to be along $\hat{z}$,  a post-measured ensemble of a product state $\ket{00}$, and a maximally entangled state $(\ket{01}+\ket{10})/\sqrt{2}$ is created, where these states occur with probabilities $1/3$ and $2/3$ respectively. Therefore, the protocol for localizing maximum resource in the form of  entanglement over any two qubits in the case of three-qubit GHZ and W states can also be viewed as a probabilistic protocol for creating maximally entangled states.  Since six rounds of noisy POVMs can provide $\mathcal{E}_{\hat{m}_3^{\text{opt}}}$ (see Sec.~\ref{subsec:gGHZ}), it is logical to ask whether the post-measured ensemble on the qubits $1$ and $2$ hosts maximally entangled states too. To investigate this, we define the fidelity of a member $\varrho_{(\boldsymbol{\lambda},\vec{\boldsymbol{\eta}})}$ (see Sec.~\ref{subsec:multiple_rounds}) of the post-measured ensemble corresponding to the measurement outcome $\boldsymbol{\lambda}$ and the optimum measurement direction $\hat{\mathbf{u}}$ with $\vec{\boldsymbol{\eta}}=\boldsymbol{\eta}\otimes\hat{\mathbf{u}}$ as 
\begin{eqnarray}
    \mathcal{F}_{(\boldsymbol{\lambda},\vec{\boldsymbol{\eta}})}&=&\max_{(U_{1},U_{2})}\bra{\psi}(U_{1}^{\dagger}\otimes U_{2}^{\dagger})\varrho_{(\boldsymbol{\lambda},\vec{\boldsymbol{\eta}})}(U_{1}\otimes U_{2})\ket{\psi}.
    \label{eq:fidelity}
\end{eqnarray}
Here, $\{U_{1}, U_{2}\}$ are single-qubit local unitary operators,  and  $\ket{\psi}$ is a two-qubit maximally entangled state. For simplicity, similar to  Secs.~\ref{subsec:gGHZ} and \ref{subsec:ghz_class}, we assume $\eta_{(3,r)}=\eta$ $\forall$ $r=1,2,\cdots,6$, and use the forms of $\hat{\mathbf{u}}$ given in Eqs.~(\ref{eq:optimal_mm_pattern_gghz}) and (\ref{eq:optimal_mm_pattern_w_2}) depending on whether the LE is computed on a three-qubit GHZ, or a W state.

\begin{figure}
\includegraphics [width=\linewidth]{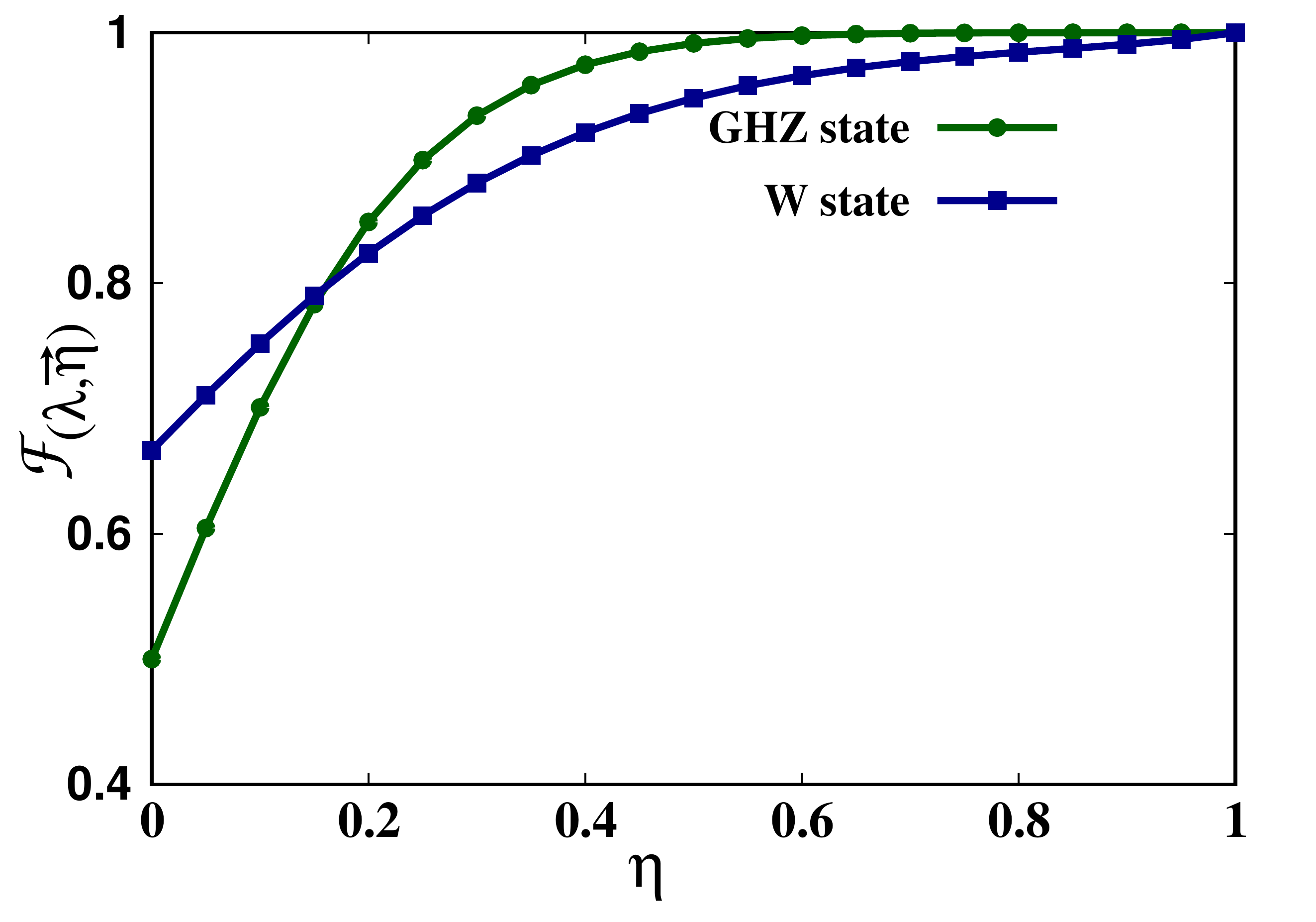}
\caption{$\mathcal{F}_{(\boldsymbol{\lambda},\vec{\boldsymbol{\eta}})}$ with the variation of $\eta$, where the measurement outcome is given by $\boldsymbol{\lambda}=\begin{bmatrix}
    +1 & +1 & +1 & +1 & +1 & +1
\end{bmatrix}$ for GHZ states and  $\boldsymbol{\lambda}=\begin{bmatrix}
    +1 & +1 & +1 & +1 
\end{bmatrix}$ for the W states. Both the axes are dimensionless.} 
\label{fig:ghz_w_fidelity}
\end{figure}

In the case of three-qubit GHZ state,   when six rounds of noisy measurements are performed, $81.25\%$ of the states in the post-measured ensemble corresponding to a MM $\hat{\mathbf{u}}$ given in Eq.~(\ref{eq:optimal_mm_pattern_gghz}) have a fidelity $\mathcal{F}_{(\boldsymbol{\lambda},\vec{\boldsymbol{\eta}})}>0.99$, while for the rest of the states, $0.9\leq \mathcal{F}_{(\boldsymbol{\lambda},\vec{\boldsymbol{\eta}})}\leq 0.92$ with \(\eta = 0.8\). Note, however, that the fidelity and the percentage mentioned above depends on the rounds of measurements performed on the third qubit. For example,  in the next round, i.e., when \(R=7\), keeping the same value of $\eta$, we find that $90.63\%$ of the states  have a fidelity $\mathcal{F}_{(\boldsymbol{\lambda},\vec{\boldsymbol{\eta}})}>0.98$, while for the rest of the states, $0.56\leq \mathcal{F}_{(\boldsymbol{\lambda},\vec{\boldsymbol{\eta}})}\leq 0.57$.

On the other hand, in the case of the $W$ state, with $R=4$,  $\mathcal{F}_{(\boldsymbol{\lambda},\vec{\boldsymbol{\eta}})}>0.95$ for $50\%$ states of the post-measured ensemble corresponding to the $\hat{\mathbf{u}}$ given in Eq.~(\ref{eq:optimal_mm_pattern_w_1}), while for the rest $50\%$ of states, $0.5\leq \mathcal{F}_{(\boldsymbol{\lambda},\vec{\boldsymbol{\eta}})}\leq 0.66$. However, similar to the GHZ state, the structure of the post-measured ensemble depends on the number of rounds of measurements. In Fig.~\ref{fig:ghz_w_fidelity}, we plot the variation of  $\mathcal{F}_{(\boldsymbol{\lambda},\vec{\boldsymbol{\eta}})}$ as a function of $\eta$ in the case of GHZ and W states, corresponding to a specific measurement outcome given by $\boldsymbol{\lambda}=\begin{bmatrix}
    +1 & +1 & +1 & +1 & +1 & +1
\end{bmatrix}$ and $\boldsymbol{\lambda}=
\begin{bmatrix}
     +1 & +1 & +1 & +1 
\end{bmatrix}$.

It is worthwhile to note that in the case of the W state, all post-measured states corresponding to a measurement outcome $\boldsymbol{\lambda}$ with $\lambda_{(3,1)}=+1$ has a fidelity $> 0.95$ with a maximally entangled state. Therefore, for using noisy measurements to create highly entangled two states that can be used as resource in specific quantum protocols, starting from a three-qubit W state, one requires $\eta>0.5$, and $\lambda_{(3,1)}=+1$, which helps in determining whether one would continue measuring on the third qubit after the first round of measurement.  It also establishes that not only average entanglement can be achieved via sequential measurements, the desired highly resourceful states can be prepared with the help of sequential noisy measurements, thereby overcoming the destructive influence of noise on quantum processes.

\begin{figure}
\includegraphics [width=\linewidth]{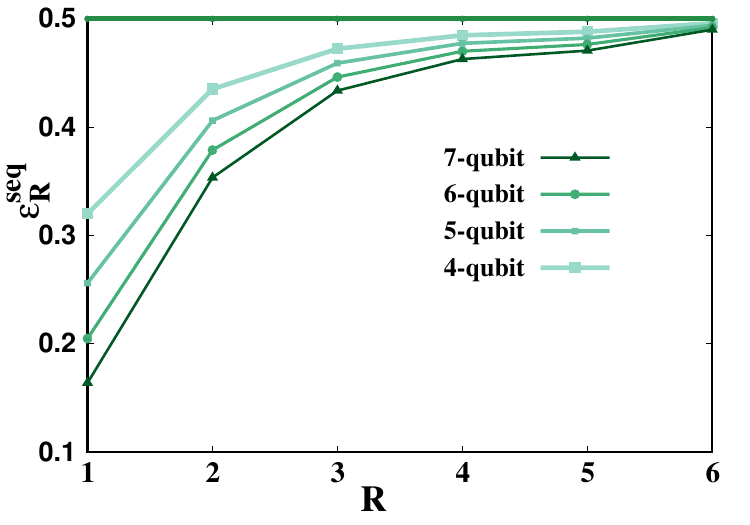}
\caption{$\mathcal{E}_{R}^{\text{seq}}$ against $R$ with $1\leq R\leq 6$. All the plots correspond to multiparty \text{GHZ} states with the number of qubits shown in each plot. Here, \(\eta=0.8\) is taken in all the rounds. All quantities plotted are dimensionless.} 
\label{fig:ghz_multiparty}
\end{figure}

\section{Multiple measurements on multiqubit states}
\label{sec:multiple_qubits}

We now go beyond three-qubit systems, and investigate whether sequential unsharp measurements are beneficial in localizing entanglement over any two chosen qubits, when measurements are performed on all of the rest of the qubits.

\noindent\textbf{Generalized GHZ state of $N$ qubits.} The \(N\)-qubit gGHZ state reads as $\ket{\text{gGHZ}}_{N}=c_{0}\ket{0}^{\otimes N}+c_{1}\ket{1}^{\otimes N}$, 
for which sharp projection measurements along $\hat{x}$ on $N-2$ qubits creates an ensemble of states of the form (\ref{eq:ghz_state_ex}) on the rest of the two qubits, leading to an LE $|c_0c_1|$. However, in the case of noisy measurements of multiple rounds on multiple qubits, the calculation of LE to its full generality and for arbitrary $\boldsymbol{\eta}$ is difficult. To probe this, we assume $\eta_{(b,r)}=\eta$ $\forall$ $b$ and $r$, and   numerically explore the $N$-qubit gGHZ states with moderate $N$ ($3\leq N\leq 7$).  Irrespective of the value of $\eta$, we find that 
\begin{enumerate}
 \item $\mathcal{E}_{\hat{m}_3^{\text{opt}}}-\mathcal{E}_{R}^{\text{seq}}\leq 10^{-2}$ for $R=6$, and 
    \item two of the \emph{optimum} $(N-2)\times R$ MMs $\hat{\mathbf{u}}$ among the OPS $\mathcal{S}_\perp$,  which is currently a set of
    all possible $\hat{\mathbf{u}}$ such that  (see Sec.~\ref{subsec:gGHZ} and Eq.~(\ref{eq:pattern_single_qubit})) 
    \begin{eqnarray}
    \hat{u}_{(b,r)}&\in&\{\hat{x},\hat{y},\hat{z}\},
    \hat{u}_{(b,r-1)}\perp\hat{u}_{(b,r)}\perp\hat{u}_{(b,r+1)}
    \label{eq:pattern_multiple_qubit}
    \end{eqnarray}
    for all $b$ with $1\leq r\leq R=6$, to be always 
\begin{eqnarray}
    \hat{\mathbf{u}}&=&
    \begin{bmatrix}
       \hat{x} & \hat{y} & \hat{x} & \hat{y} & \hat{x} & \hat{y}  \\
       \hat{y} & \hat{x} & \hat{y} & \hat{x} & \hat{y} & \hat{x}  \\
       \hat{x} & \hat{y} & \hat{x} & \hat{y} & \hat{x} & \hat{y} \\
       \hat{y} & \hat{x} & \hat{y} & \hat{x} & \hat{y} & \hat{x} \\       
       \vdots & \vdots & \vdots & \vdots & \vdots & \vdots    
    \end{bmatrix},
\end{eqnarray}
and
\begin{eqnarray}
    \hat{\mathbf{u}}&=&
    \begin{bmatrix}
       \hat{x} & \hat{y} & \hat{x} & \hat{y} & \hat{x} & \hat{y} \\
       \hat{x} & \hat{y} & \hat{x} & \hat{y} & \hat{x} & \hat{y} \\
       \hat{x} & \hat{y} & \hat{x} & \hat{y} & \hat{x} & \hat{y}\\
       \hat{x} & \hat{y} & \hat{x} & \hat{y} & \hat{x} & \hat{y} \\     
       \vdots & \vdots & \vdots & \vdots & \vdots & \vdots  
    \end{bmatrix},
\end{eqnarray}
both satisfying Eq.~(\ref{eq:pattern_multiple_qubit}), where $(N-2)$ can be even, or odd. 
\end{enumerate}
In Fig.~\ref{fig:ghz_multiparty}, we plot $\mathcal{E}_R^{\text{seq}}$ against $R$, $1\leq R\leq 6$, and observe that the value of $\mathcal{E}_{R}^{\text{seq}}$  increases monotonically with $R$, and approaches $\mathcal{E}_{\hat{m}^{\text{opt}}}$. Some of the important observations are as follows: $(I)$ $\mathcal{E}_1^{\text{seq}}$ decreases with increasing $N$ (see table \ref{tab:rounds}). $(II)$ The above result may give the impression of the requirement of a larger value of $R$ for obtaining $\mathcal{E}_{\hat{m}_3^{\text{opt}}}$ in the case of a higher value of $N$. Our analysis suggests that the number of rounds of noisy measurements required for obtaining $\mathcal{E}_{\hat{m}_3^{\text{opt}}}$, up to an error of $5\times 10^{-3}$, indeed increases slowly with $N$.  See, for example, Table~\ref{tab:rounds} for a demonstration in the case of the $N$-qubit GHZ state for $3\leq N\leq 7$, where $6$ rounds of noisy measurements have already achieved $\mathcal{E}_{\hat{m}_3^{\text{opt}}}$ up to an error of $10^{-2}$.  
\begin{table}
\begin{tabular}{|c|c|c|}
\hline 
$N$  & $\mathcal{E}_1^{\text{seq}}$ & $\mathcal{E}_6^{\text{seq}}$ \\ 
\hline
$3$   & 0.4 & 0.498 \\
\hline 
$4$   & 0.32 & 0.496\\
\hline 
$5$   & 0.256 & 0.494\\
\hline 
$6$   &  0.205 & 0.492\\
\hline
$7$   &  0.164 & 0.490\\
\hline 
\end{tabular}
\caption{Values of $\mathcal{E}_1^{\text{seq}}$, $\mathcal{E}_6^{\text{seq}}$ for $N$-qubit GHZ state with moderate values of $N$, for which $\mathcal{E}_{\hat{m}^{\text{opt}}}=0.5$.}
\label{tab:rounds}
\end{table}

\begin{figure}
\centering 
\includegraphics[width=\linewidth]{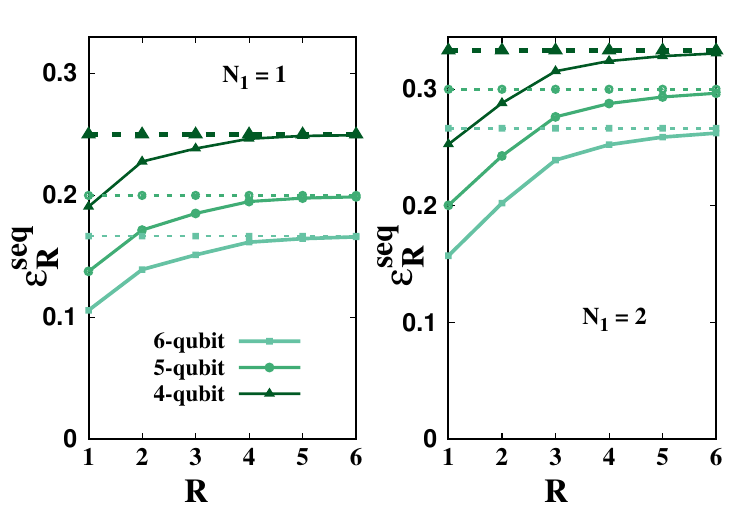}
\caption{Variations of $\mathcal{E}_R^{\text{seq}}$ as a function of $R$ ($1\leq R\leq 6$) for Dicke states with $N=4,5,6$, and  $N_1=1,2$. All quantities plotted are dimensionless.} 
\label{fig:d1_d2_multiparty}
\end{figure}

\noindent\textbf{Dicke states of $N$ qubits.} As observed in case of gW and W class states, the rounds required to achieve $\mathcal{E}_{\hat{m}^{\text{opt}}}$ is lower compared to the GHZ class states or in general Haar uniformly generated states. Let us check whether such trends remain same for a multipartite generalization of W class states or not. The class of symmetric states that remain invariant under permutation of parties are known as Dicke states of $N$ qubits, where $N_{0}$ qubits are in the ground states $\ket{0}$, and the rest of the $N_{1}=N-N_{0}$ qubits are in the excited states $\ket{1}$. A Dicke state with $N_{1}$ excited qubits can be written as  
\begin{eqnarray}
    \ket{D(N,N_{1})}&=& \frac{1}{\sqrt{\genfrac{()}{}{0pt}{}{N}{N_1} }}\sum_{i}\mathcal{P}_{i}(\ket{0}^{\otimes N-N_{1}}\ket{1}^{\otimes N_{1}}),
    \label{eq:dicke_state}
\end{eqnarray} 
with $N_{1}=1,2,...,N$ for a fixed $N$. Here, \(\{\mathcal{P}_{i}\}\) is the set of all possible permutations of $N_{0}(N_{1})$ qubits at the ground (excited) state, such that $N_{0}+N_{1}=N$. Note that the complexity in computation of $\mathcal{E}_R^{\text{seq}}$  increases significantly with increasing $N$, since one has to deal with $2^{R(N-2)}$ measurement outcomes for $R$ rounds of measurements on each of the $N-2$ qubits. We perform numerical analysis for moderate values of $N$ ($N\leq 6$) and find that 
\begin{enumerate}
   \item similar to the $N$-qubit gGHZ states, $\mathcal{E}_{\hat{m}_3^{\text{opt}}}-\mathcal{E}_{R}^{\text{seq}}\leq 5\times 10^{-3}$ for $R=6$, and
    \item one can always find an \emph{optimum} $(N-2)\times R$ MMs $\hat{\mathbf{u}}$ belonging OPS $\mathcal{S}_\perp$ such that for $N_1=N/2$ (with $N$ even), and $N_1=(N\pm 1)/2$ (with $N$ odd), 
    \begin{eqnarray}
    \hat{\mathbf{u}}&=&
    \begin{bmatrix}
       \hat{z} & \hat{x} & \hat{z} & \hat{x} & \hat{z} & \hat{x} \\
       \hat{z} & \hat{x} & \hat{z} & \hat{x} & \hat{z} & \hat{x} \\
       \hat{z} & \hat{x} & \hat{z} & \hat{x} & \hat{z} & \hat{x} \\
       \hat{z} & \hat{x} & \hat{z} & \hat{x} & \hat{z} & \hat{x} \\      
       \vdots & \vdots & \vdots & \vdots & \vdots & \vdots
    \end{bmatrix},
    \label{eq:pattern_pattern_dicke_1}
\end{eqnarray}
while for all other values of $N_1$ corresponding to a fixed value of $N$, 
    \begin{eqnarray}
    \hat{\mathbf{u}}&=&
    \begin{bmatrix}
       \hat{x} & \hat{y} & \hat{z} & \hat{x} & \hat{y} & \hat{z} \\
       \hat{y} & \hat{x} & \hat{z} & \hat{y} & \hat{x} & \hat{z} \\
       \hat{x} & \hat{y} & \hat{z} & \hat{x} & \hat{y} & \hat{z} \\
       \hat{y} & \hat{x} & \hat{z} & \hat{y} & \hat{x} & \hat{z} \\      
       \vdots & \vdots & \vdots & \vdots & \vdots & \vdots
       \label{eq:pattern_pattern_dicke_2}       
    \end{bmatrix},
\end{eqnarray}
where $N-2$ can be even, or odd. Note that Eq.~(\ref{eq:pattern_multiple_qubit}) is satisfied for these matrices also. The forms for $\hat{\mathbf{u}}$ given in both Eqs.~(\ref{eq:pattern_pattern_dicke_1}) and (\ref{eq:pattern_pattern_dicke_2}) satisfy Eq.~(\ref{eq:pattern_multiple_qubit}). In Fig.~\ref{fig:d1_d2_multiparty}, we plot the variations of $\mathcal{E}_R^{\text{seq}}$ with $R$. In this case also, we observed that with the increase of assisting qubits, higher number of rounds are essential to reach $\mathcal{E}_{\hat{m}^{\text{opt}}}$.
\end{enumerate}

\section{Conclusion}
\label{sec:conclude}

Quantum protocols often require the preparation of specific resourceful (entangled) states among a few nodes, of a network which can subsequently enable the desired  quantum information processing scheme. Approximate measurements are often used to prepare these states, or a given resource over a chosen subsystem in a probabilistic approach, which may be required to execute certain quantum information scheme. However, the measurement apparatus inevitably interact with the environment, thereby performing noisy (unsharp) measurements. While the impact of noise on states has been extensively studied, the effect of noisy measurements on such resource-localization protocols remain unexplored.

In this work, we focused on localizing entanglement over a chosen subsystem via performing \emph{noisy} measurements over the rest of the subsystems consisting of assisting qubits. Since noisy measurements do not fully destroy the entanglement between the measured and the unmeasured parties, it allows one to perform multiple rounds of noisy measurements on the same qubit. We formulated entanglement concentration protocol with sequential noisy measurements. We demonstrated, for a set of paradigmatic pure states, that multiple rounds of noisy measurements  can achieve the entanglement localizable  via sharp projection measurements. We also identified specific patterns in the optimal measurement directions corresponding to the sequence of different measurement rounds, and  found the possibility of using consecutive noisy measurements as a protocol for preparing desired quantum states. We showed that  three-qubit W class states and multiqubit Dicke states have less impacts of noisy measurements on  entanglement localization compared to arbitrary multiqubit states.

Our study reveals two intriguing characteristics that can be helpful even in cases where the measurement operators are noisy --  (1) the potential for sequential measurements, which has been employed in the literature for different purposes \cite{Mal16,Silva15,asmita19,Colbeck20,Halder2022, Bera18,Srivastava21, Anwer2021, sasmal18, Shenoy19, ROY2021127143, Das2023}, in noise mitigation, and (2) the possibility of partial destruction of links between assisting and non-assisting parties providing more control to the assisting parties, which could be important in the design of secure communication.   

\acknowledgements
 
SM, PH, and ASD acknowledge the support from Interdisciplinary Cyber Physical Systems (ICPS) program of the Department of Science and Technology (DST), India, Grant No.: DST/ICPS/QuST/Theme- 1/2019/23. AKP and ASD 
acknowledge the support from the Anusandhan National Research Foundation (ANRF) of the Department of Science and Technology (DST), India, through the Core Research Grant (CRG) (File No. CRG/2023/001217, Sanction Date 16 May 2024). The authors acknowledge the use of \href{https://github.com/titaschanda/QIClib}{QIClib} -- a modern C++ library for general purpose quantum information processing and quantum computing (\url{https://titaschanda.github.io/QIClib}), and the cluster computing facility at Harish-Chandra Research Institute.

\bibliography{ref}

\end{document}